\documentclass[aps,twocolumn,superscriptaddress]{revtex4} %pour twolcolumn
\usepackage[T1]{fontenc}
\usepackage[dvips]{graphicx}
\usepackage[dvips]{color}
\usepackage{caption}
\usepackage{txfonts}
\usepackage{natbib}
\newcommand{\degreeC}{\ensuremath{^\circ}C}

\begin{document}

\title{Domain structure sequence in ferroelectric Pb(Zr$_{0.2}$Ti$_{0.8}$)O$_3$ thin film on MgO}
\author{Pierre-Eymeric Janolin}
\date{\today}
\affiliation{%
Laboratoire Structures, Propri\'{e}t\'{e}s et Mod\'{e}lisation des Solides, CNRS UMR 8580, \'{E}cole Centrale Paris, 92295 Ch\^{a}tenay-Malabry Cedex, France}
\author{Fran\c coise Le Marrec}
\affiliation{%
Laboratoire de Physique de la Mati\`{e}re Condens\'{e}e, Universit\'{e} de Picardie Jules Verne, 80039 Amiens, France}
\author{Erling Ringgaard}
\affiliation{%
Ferroperm Piezoceramics A/S, Hejreskovvej 18A DK-3490 Kvistg\aa rd, Denmark}
\author{Bernard Fraisse}
\affiliation{%
Laboratoire Structures, Propri\'{e}t\'{e}s et Mod\'{e}lisation des Solides, CNRS UMR 8580, \'{E}cole Centrale Paris, 92295 Ch\^{a}tenay-Malabry Cedex, France}
\author{Brahim Dkhil}
\affiliation{%
Laboratoire Structures, Propri\'{e}t\'{e}s et Mod\'{e}lisation des Solides, CNRS UMR 8580, \'{E}cole Centrale Paris, 92295 Ch\^{a}tenay-Malabry Cedex, France}
\begin{abstract}
The structural evolution of a polydomain ferroelectric Pb(Zr$_{0.2}$Ti$_{0.8}$)O$_3$ film was studied by temperature-dependent X-ray diffraction. Two critical temperatures were evidenced: T*=740K, corresponding to a change in the domain structure (a/c/a/c to a1/a2/a1/a2), and T$_c^{film}$=825K where the film undergoes a ferroelectric-paraelectric phase transition. The films remains tetragonal on the whole range of temperature investigated. The evolutions of the domain structure and lattice parameters were found to be in very good agreement with the calculated domain stability map and theoretical temperature-misfit strain phase diagram respectively. 
\end{abstract}

\maketitle

Among the ferroelectric materials, Pb(Zr$_{1-x}$Ti$_x$)O$_3$ (PZT 1-x/x) is used in most of the optical and microelectronic devices because of its exceptional properties. With temperature this solid solution undergoes a paraelectric (cubic) to ferroelectric (slightly distorted unit cell) phase transition. For Ti-rich compositions, the symmetry of the ferroelectric phase is tetragonal whereas the Zr-rich composition is rhombohedral (except for a narrow region close to PbZrO$_3$). The two phases are separated by a morphotropic phase boundary (MPB) at x $\approx$ 0.48 which symmetry at low temperature has been extensively discussed \cite{Noheda1999,Noheda2000,Lima2001,Vanderbilt2001,Frantti2002,SouzaFilho2002,Noheda2002,Hatch2002,Kornev2006}. 
Because of the anisotropy of their properties, controlling the domain structure of ferroelectric materials is desirable; this can be achieved through epitaxial growth of thin films. 

An extension of the tetragonal phase toward the MPB was observed\cite{Oh1998} and different domain structures have been observed for tetragonal PZT films deposited on MgO: epitaxial PZT 52/48 has been deposited by different techniques including sol-gel\cite{Nashimoto1995}, pulsed laser deposition (PLD)\cite{Lee1992} and sputtering\cite{Oh1998}. For the three techniques a 100\% \textit{c}-domain structure has been reported with thickness ranging from 90~nm to 700~nm. PZT 40/60 has also been deposited by PLD\cite{Kim1999} and the domain structure was found to be 100\% \textit{c}-domains. A polydomain structure has been reported by Lee \textit{et al.}\cite{Lee1999_b}, with an decrease of the volume fraction of \textit{c}-domains ($\alpha^c$) with increasing Ti concentration from 90\% for PZT 32/68 down to 75\% for pure PbTiO$_3$. PbTiO$_3$, the end-member of PZT, has been studied as a thin film deposited on MgO by many authors. 
$\alpha^{c}$ was reported to vary from 57 to 90\%, depending on the thickness of the film and to increase\cite{Lee2000} as well as to decrease\cite{Kwak1994} with film thickness. It should be underlined that as these volume fractions have been calculated in different ways, the comparison is not always straightforward (see \cite{Hsu1995}). In addition Roemer \textit{et al.} showed that $\alpha^{c}$ is strongly dependent on the deposition temperature and partial pressure of O$_2$ for thin film pulsed laser deposited\cite{Roemer2004}. The reduction of $\alpha^{c}$ with temperature was predicted\cite{Pertsev1996,Alpay1998} and measured for PbTiO$_3$ films deposited on MgO\cite{Kim1996,Batzer1996,Lee1999_a,Lee2000}. 

Lattice parameters evolution with temperature for PbTiO$_3$ thin films deposited on MgO was reported by Kim \textit{et al.}\cite{Kim1996} and Batzer \textit{et al.}\cite{Batzer1996}.
The transition temperature for these films have been reported to exhibit different behavior and remains unclear: Batzer \textit{et al.}\cite{Batzer1996} reported the transition temperature of the film, T$_C^{film}$, to be lowered by 10~K (resp. 35~K) when heating (resp. cooling) as a consequence of the compressive stress arising from the difference in the thermal expansion coefficients between the film and the substrate. Kim \textit{et al.}\cite{Kim1996} reported T$_C^{film}$ to be 10~K higher (resp. 40~K lower) than the bulk when heating (resp. cooling). 
Temperature evolution of Pb(Zr$_{1-x}$Ti$_{x}$)O$_3$ ($x$=1,0.92,0.84,0.76,0.68) thin films deposited on MgO was reported by Lee \textit{et al.}\cite{Lee1999_b}. The out-of-plane lattice parameters as well as T$_C^{film}$ were equal to the corresponding bulk values, i.e. the films considered were strain-free.

We have performed X-ray diffraction measurements to follow the evolution with temperature from 14~K up to 925~K of both the lattice parameters and fraction of \textit{c}-domains, $\alpha^c$, of PZT 20/80 films deposited on MgO(001). These evolutions allow direct comparison with the calculations done on multidomain PZT films as well as the discussion of the symmetry of the paraelectric phase.
At room temperature, bulk PZT is tetragonal with a$_b$=3.950 \AA\ and c$_b$=4.150 \AA\ and c$_{MgO}$=4.213 \AA. 
100 nm PZT 20/80 film was deposited by PLD with a KrF excimer laser ($\lambda$=248 nm), under 0.2 mbar O$_2$ with a laser repetition rate of 2 Hz at a fixed fluence of 1.6 J/cm$^2$. Ceramic Pb$_{1.1}$(Zr$_{0.2}$Ti$_{0.8}$)O$_3$ target was used to deposit the 100 nm film at a substrate temperature of 650 \degreeC.
The precise thickness of a PZT 20/80 film deposited on SrTiO$_3$ was determined through finite size oscillations and the thickness of the PZT 20/80 film deposited on MgO was estimated from the calculated deposition rate. The composition of the PZT film have been checked through volume calculation\cite{Frantti2002}.
Lattice parameters determination was carried out on a high-precision diffractometer using Cu-K$_\beta$ wavelength issued from a 18kW rotating anode generator. 

At room temperature, the out-of-plane lattice parameters of the films were calculated from (00l)$_{l=2,3,4}$ Bragg reflections to improve accuracy and to correct any misalignment of the sample and are a$_f^{\perp}$=3.974 \AA\ and c$_f^{\perp}$=4.129 \AA .
It has not been possible to determine the in-plane lattice parameters even though the out-of-plane ones give a good  approximation of their value\cite{Lee2001}. The film is polydomain (a/c/a/c) and, from the intensity ratio of the (002) and (200) PZT 20/80 rocking curves, composed of $\alpha^c\approx$75\% of \textit{c}-domains and $\approx$25\% of \textit{a}-domains. These values are typical for PbTiO$_3$ films deposited on MgO\cite{Kwak1994,Lee2000} but somewhat smaller than the interpolated value from \cite{Lee1999_b} which is consistent with our film being not strain-free. 

Intriguingly, the film is mainly composed of \textit{c}-domains even though it experiences a tensile stress at room temperature. This tensile stress results from firstly the tensile stress developed at the deposition temperature because of both the lattice mismatch between PZT and MgO and the deposition process itself, and secondly from the intrinsic stress associated to the phase transition. This tensile stress is only partially compensated by the compressive thermal stress ($\alpha_{MgO}$=12.4$\cdot$10$^{-6}$~K$^{-1}$ and $\alpha_{PZT}$=9.2$\cdot$10$^{-6}$~K$^{-1}$) experienced when cooling down from T$_d$ to RT.

\begin{figure}[htb]
	\centering
		\includegraphics[width=0.4\textwidth]{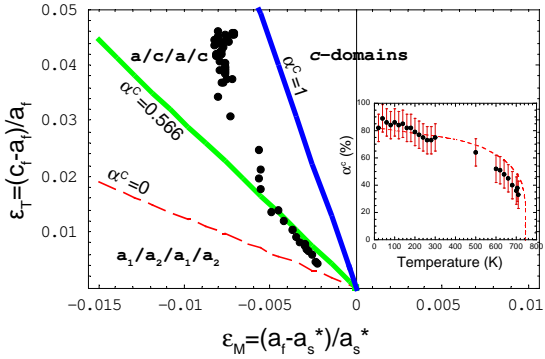}
			\caption{Calculated domain stability map for PZT(20/80)/MgO. The solid circles are the experimental points from T=14~K to 740~K from top to bottom. For each region, delimited by thick solid lines, the most stable phase is indicated. The a/c/a/c phase is metastable in the region between the lines $\alpha^c$=0.566 (solid) and $\alpha^{c}$=0 (dashed). Inset: $\alpha^c$ versus temperature with the fit using $\alpha^c=\frac{1-0.5\cdot\eta}{1-\eta}\frac{1+\nu}{1-\eta}\frac{\epsilon_M}{\epsilon_T}$ with $\eta$=0.1 and $\nu$=0.36}
	\label{fig:stabilitymap}
\end{figure}

We have calculated the domain stability map for PZT(20/80) on MgO(001) following Alpay\cite{Alpay1998}.  
The experimental points have been plotted on the calculated domain stability map (see Fig.\ref{fig:stabilitymap}) and it confirms the X-ray observations: a film being slightly under tension (i.e. $\epsilon_M$\textless 0) is composed of a majority of \textit{c}-domains as long as the tetragonality ($\epsilon_T$) is high enough. Here $\epsilon_T$=(c$_f$-a$_f$)/a$_f$ is equal to 0.039 at room temperature.
 
\begin{figure}[tp]
	\centering
		\includegraphics[width=0.4\textwidth]{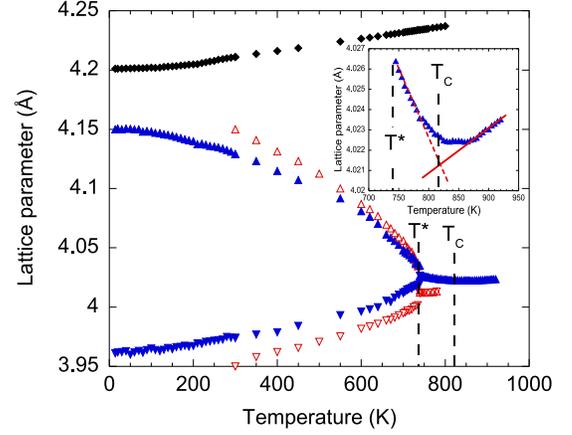}
	\caption{Lattice parameters of PZT20/80 bulk (c$_b$ \textcolor{red}{$\triangle$}, a$_b$ \textcolor{red}{$\triangledown$}) and thin films (c$^{\perp}_f$ \textcolor{blue}{$\blacktriangle$}, a$^{\perp}_f$ \textcolor{blue}{$\blacktriangledown$}) deposited on MgO ($\vardiamondsuit$). Inset: High-temperature evolution of the out-of-plane lattice parameter. Full line arises from a calculation based on an elastic behavior of the film (see text).}
	\label{fig:PZTsurMgO}
\end{figure}

For 1\textgreater $\alpha^c$\textgreater 0.566 (i.e. 14\textless T(K)\textless 600~K) the observed (a/c/a/c) domain structure is in agreement with the calculated domain stability map. When $\alpha^c$ becomes smaller than 0.566 (which corresponds to T\textgreater600~K) the a/c/a/c phase becomes metastable and disappears for $\alpha^{c}$=0, at T*. Therefore, from 600~K to 740~K the a/c/a/c and a$_1$/a$_2$/a$_1$/a$_2$ phases should coexist.
The temperature evolution of $\alpha^{c}$ has also been determined (see inset Fig.\ref{fig:stabilitymap}). The decrease of $\alpha^c$ with temperature has been predicted by Pertsev \textit{et al.}\cite{Pertsev1996} and is consistent with the evolution of the experimental points toward the $\alpha^{c}$=0 line (see Fig.\ref{fig:stabilitymap}) with increasing temperature (i.e. decreasing $\epsilon_T$). This decrease toward zero indicates that \textit{c}-domains nucleate for T=T$_C$ and then grow continuously as T decreases, a mechanism proposed by Kim \textit{et al.}\cite{Kim1996}. The measured $\alpha^{c}$ is however over-estimated by the fit for T\textgreater600~K. We will address this point later.

Figure \ref{fig:PZTsurMgO} shows the temperature evolution of the lattice parameters of both PZT 20/80 thin film and MgO between 14~K and 925~K. For comparison the lattice parameters of the PZT 20/80 target used for deposition are also plotted from 300~K to 800~K. Assuming the overall evolution is depicted by the out-of-plane lattice parameters, the lattice parameters of the film exhibit a bulk-like evolution with temperature, and they become equal when heating at T* close to T$_C^{bulk}$=740~K. 
However, the most intriguing feature in the evolution is the out-of-plane lattice parameter above T*. Indeed when heating, it begins to \emph{decrease} linearly with temperature for T*\textless T\textless825~K and then increases linearly (see inset Fig.\ref{fig:PZTsurMgO}) for T\textgreater825~K.

This peculiar behavior may be explained considering the following evolution, from the deposition temperature T$_d$: at T$_d$ the film is in equilibrium with an effective substrate (MgO*), \textit{i.e.} with the misfit-dislocation-modified MgO (a$^{//}_f$=a$_{MgO*}$ at T$_d$).
In the paraelectric phase, we consider the film as purely elastic with its in-plane parameter driven by the effective substrate. It is then possible to calculate the temperature evolution of the out-of-plane parameter\cite{Janolin2006}. This calculation gives a straight line (represented by the full line on the Inset of Fig.(\ref{fig:PZTsurMgO})) that fits very well the high-temperature evolution of the out-of-plane parameter. At $\sim$825~K the film cannot be considered as purely elastic anymore. The turning point in the evolution is due to a phase transition. This brings T$_C^{film}$=825~K which corresponds to the paraelectric to ferroelectric phase transition temperature. When cooling from T$_C^{film}$ to T*, the film is tetragonal and ferroelectric, and the evolution of the out-of-plane parameter is imposed by the phase transition. It becomes equal to the in-plane parameter at T*, only temperature where the film adopts a pseudo-cubic symmetry remaining however ferroelectric.

\begin{figure}[tp]
	\centering
		\includegraphics[width=0.4\textwidth]{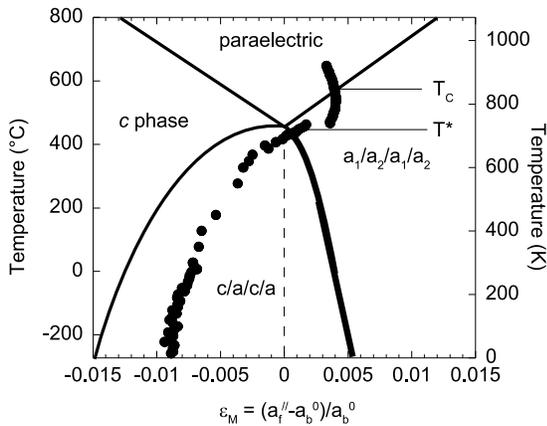}
	\caption{Experimental misfit strain experienced by the film on the temperature-misfit strain diagram calculated by Koukhar \textit{et al.}\cite{Koukhar2006}.}
	\label{fig:phdiag}
\end{figure}

This proposed evolution is supported by the temperature-misfit strain phase diagram. It is indeed possible to calculate the misfit strain experienced by the film for each temperature: $\epsilon_{M}=\frac{a_f^{//}-a_b^0}{a_b^0}$ and to compare it with Koukhar \textit{et al.}\cite{Koukhar2006}'s recent calculation of the temperature-misfit strain phase diagram for polydomain epitaxial Pb(Zr$_{1-x}$Ti$_x$)O$_3$ thin films with dense domain structure (see Fig. \ref{fig:phdiag}). 
First of all, the domain structure observed is in agreement with the temperature-misfit strain phase diagram: from 14~K up to T*, the film is polydomain and the evolution of the experimental points with temperature is in agreement with the decreasing of $\alpha^c$ on heating, as the experimental points tend toward the a$_1$/a$_2$/a$_1$/a$_2$ phase which consists of \textit{a}-domains only.
Then the experimental points cross at T* the line separating the a/c/a/c and a$_1$/a$_2$/a$_1$/a$_2$ phases, confirming the ferroelectric nature of the phase above T*. Moreover, the a/c/a/c $\rightarrow$ a$_1$/a$_2$/a$_1$/a$_2$ transition is of first order, according to the phase diagram. This explains the apparent under-estimation of $\alpha^c$ (see inset Fig.(\ref{fig:PZTsurMgO})) when compared to the fit: the measured volume fraction of \textit{c}-domains in the temperature range where the two structures coexist takes into account not only the \textit{a}-domains of the a/c/a/c structure but also the \textit{a}-domains of the a$_1$/a$_2$/a$_1$/a$_2$ structure, resulting in an over-estimation of the \textit{a}-domains and hence an under-estimation of the \textit{c}-domains. In other words, when the two structures coexist (600~K-740~K), we measured $\alpha^c$ for a (a/c/a/c + a$_1$/a$_2$/a$_1$/a$_2$) structure, which is necessarily smaller than for the a/c/a/c structure alone.
Finally, the phase (or structure) sequence proposed and the corresponding transition temperatures are in very good agreement with the temperature-misfit phase diagram. It is worth noting that by shifting the calculated phase diagram toward more positive misfit strains we would get a better fit to our experimental data. This shift would cause the phase diagram to be not centered anymore with respect to the zero misfit value\cite{Lai2005}.

In conclusion, we have shown that, despite its domain structure, a polydomain ferroelectric thin film is not necessarily strain-free and exhibits a complex evolution with temperature: a change in the domain structure takes place at T*, close to T$_C^{bulk}$ from a polydomain a/c/a/c phase to a a$_1$/a$_2$/a$_1$/a$_2$ phase. The Curie temperature of the film is shifted toward higher temperature by $\sim$60~K and the symmetry of the lattice remains tetragonal from 14~K up to temperatures higher than the deposition temperature. This results are in full agreement with both the domain stability map we have calculated and with the temperature-misfit strain phase diagram.

\end{document}